\documentclass[aps,prd,onecolumn,groupedaddress,showpacs,nofootinbib,amssymb]{revtex4}
\usepackage[dvips]{graphicx}
\usepackage{amssymb}
\usepackage{amsmath}
\usepackage{graphicx,,color}
\usepackage{amsfonts}
\usepackage{bm}
\usepackage{cancel}
\usepackage{comment}

\newcommand\be{\begin{equation}}
\newcommand\ee{\end{equation}}

\allowdisplaybreaks[4]

\begin{document}

\title{Viable Inflation in Scalar-Gauss-Bonnet Gravity and Reconstruction from Observational Indices}
\author{S.D. Odintsov,$^{1,2,3}$\,\thanks{odintsov@ieec.uab.es}
V.K. Oikonomou,$^{4,5,3}$\,\thanks{v.k.oikonomou1979@gmail.com}}
\affiliation{$^{1)}$ ICREA, Passeig Luis Companys, 23, 08010 Barcelona, Spain\\
$^{2)}$ Institute of Space Sciences (IEEC-CSIC) C. Can Magrans s/n,
08193 Barcelona, Spain\\
$^{3)}$ Tomsk State Pedagogical University, 634061 Tomsk, Russia\\
$^{4)}$Department of Physics, Aristotle University of Thessaloniki, Thessaloniki 54124, Greece\\
$^{5)}$ Laboratory for Theoretical Cosmology, Tomsk State University
of Control Systems
and Radioelectronics, 634050 Tomsk, Russia (TUSUR)\\
}

\tolerance=5000

\begin{abstract}
In this paper the focus is on inflationary dynamics in the context of Einstein Gauss-Bonnet gravitational theories. We investigate the implications of the slow-roll condition on the slow-roll indices and we investigate how the inflationary dynamical evolution is affected by the presence of the Gauss-Bonnet coupling to the scalar field. For exemplification of our analysis we investigate how the dynamics of inflationary cubic, quartic order and also exponential scalar potentials are affected by the non-trivial Gauss-Bonnet coupling to the scalar field. As we demonstrate it is possible to obtain a viable phenomenology compatible with the observational data, although the canonical scalar field theory with cubic and quartic order potentials does not yield phenomenologically acceptable results. In addition, with regard to the exponential potential example, the Einstein Gauss-Bonnet extension of the single canonical scalar field model has an inherent mechanism that can trigger the graceful exit from inflation. Furthermore we introduce a bottom-up reconstruction technique, in the context of which by fixing the tensor-to-scalar ratio and the Hubble rate as a function of the $e$-foldings number, one is capable of reproducing the Einstein Gauss-Bonnet theory which generates the aforementioned quantities. We illustrate how the method works by using some relatively simple examples.
\end{abstract}


\maketitle

\section{Introduction}

One of the mysteries of the primordial evolution of our Universe is related to the inflationary era. The theoretical mechanism called inflation was initially introduced in the early 80's \cite{Guth:1980zm,Starobinsky:1982ee,Linde:1983gd} in order to solve the homogeneity, isotropy, flatness and horizon problems that haunted the standard Big Bang model of cosmology. The standard inflationary paradigm is realized by using a canonical scalar field dubbed inflaton, the quantum fluctuations of which may explain the Cosmic Microwave Background anisotropies. With regard to the latter, the latest Planck  observations \cite{Ade:2015lrj} have constrained the power spectrum of the primordial curvature perturbations to be nearly scale invariant. Up to date, quite many inflationary models have been introduced in the literature, some of which remain viable to a great extent even after the Planck observations \cite{Ade:2015lrj}, for reviews see \cite{inflation1,inflation2,inflation4,reviews1,reviews2}. We should note that the inflationary paradigm is not the only mechanism which may generate a nearly scale invariant power spectrum, since the latter can be achieved even in the context of bouncing cosmology \cite{bounce}. In this paper we shall confine ourselves on the study of slow-roll inflationary models which are string inspired, namely an Einstein Gauss-Bonnet inflationary models. Many studies of this sort appear in the literature, and for an important stream of related papers see for example \cite{Nojiri:2006je,Calcagni:2005im,Calcagni:2006ye,Cognola:2006sp,Nojiri:2005vv,Nojiri:2005jg,Nojiri:2007te,Bamba:2014zoa,Yi:2018gse,Guo:2009uk,Guo:2010jr,Jiang:2013gza,Koh:2014bka,Koh:2016abf,Kanti:2015pda,vandeBruck:2017voa,Kanti:1998jd,Nozari:2017rta,Chakraborty:2018scm}. The study of the Einstein Gauss-Bonnet inflationary models is highly motivated from the fact that the inflationary era occurs very close to the primordial era of our Universe which is believed to be described by quantum gravity. The complete description of this primordial quantum era is not known up to date, but various aspects of string theory can provide useful insights regarding this era. Then it is highly possible that a residual imprint of the quantum Universe onto the classical inflationary Universe can be found and it may have an observable effect on the primordial power spectrum. Therefore, the low energy effective theory that results from the primordial quantum gravity theory may affect the inflationary era directly. In this respect, the Einstein Gauss-Bonnet theory is an appealing candidate theory for the inflationary era, since it is obtained by a string theory. Particularly, the underlying string theory induces a curvature correction to the classical canonical scalar field theory, which depends on the Gauss-Bonnet invariant $\mathcal{G}$. Thus, among the various modified gravities \cite{reviews1,reviews2,reviews3,reviews4,reviews5,reviews6}, the Einstein Gauss-Bonnet theory is quite interesting since it originates from string theory.

In view of these appealing features of Einstein Gauss-Bonnet theories, in this paper we shall study the inflationary era produced by these theories in the slow-roll approximation. Firstly, we shall investigate how the slow-roll condition is realized in the context of these theories, and we shall calculate explicitly the slow-roll indices of the theory at hand. We shall also study several classes of inflationary models and we will calculate the observational indices of inflation, which we shall compare with the observational data coming from Planck. As we will demonstrate, the resulting theory can be compatible with the observations for a wide range of parameters. Also the second part of this paper is devoted on studying reconstructed slow-roll models of Einstein Gauss-Bonnet gravity from the observational indices. This bottom-up approach is frequently used in the literature
\cite{Odintsov:2018ggm,Odintsov:2017fnc,Narain:2017mtu,Narain:2009fy,Fumagalli:2016sof,Gao:2017uja,Lin:2015fqa,Miranda:2017juz,Fei:2017fub,Herrera:2018mvo,Herrera:2018ker} and it provides viable inflationary models. We shall provide analytic formulas for the slow-roll indices and for the spectral index of the primordial curvature perturbations and we will investigate the viability of the resulting models in detail. An similar reconstruction scheme to the one we shall develop, was presented by A. Starobinsky in \cite{starobconf}. We should note that in the article we shall use the general formalism developed in Refs. \cite{Noh:2001ia,Hwang:2005hb,Hwang:2002fp,Kaiser:2013sna}, however we shall extend it in the slow-roll approximation, and moreover we shall investigate the behavior of specific models. The new features of this work are the development of the slow-roll inflation formalism in the context of Einstein Gauss Bonnet theories and the introduction of the bottom-up reconstruction techniques from the observational indices, developed for the aforementioned theories.

This paper is organized as follows: In section II we present the essential features of Einstein Gauss-Bonnet theory and we investigate how the slow-roll condition modifies the dynamics of inflation, in terms of the slow-roll indices. We present analytic expressions for the slow-roll indices in the slow-roll approximation and we calculate the observational indices of inflation, namely the spectral index of the primordial curvature perturbations and the tensor-to-scalar ratio. In addition, we use two illustrative examples in order to exemplify our results. In section III we introduce a bottom-up reconstruction technique in which given the tensor-to-scalar ratio and the Hubble rate, one is able to find which theory realizes the aforementioned quantities. We provide analytic formulas for the slow-roll indices and for the spectral index, and we exemplify our technique by using a characteristic example. Finally the conclusions follow in the end of the paper.

Prior starting, it is worth to describe in brief the geometric background assumed in this paper, which is a flat Friedmann-Robertson-Walker (FRW) background, with line element,
\begin{equation}
\label{metricfrw} ds^2 = - dt^2 + a(t)^2 \sum_{i=1,2,3}
\left(dx^i\right)^2\, ,
\end{equation}
where $a(t)$ is the scale factor. In addition, the metric connection is the Levi-Civita, which is torsion-less, symmetric and metric compatible.

\section{Slow-roll Einstein Gauss-Bonnet Inflationary Dynamics}

\subsection{Essential Features of Einstein Gauss-Bonnet Gravity}

In this section we shall briefly review the formalism of Einstein Gauss-Bonnet gravity, and we shall adopt the notation and presentation of Ref. \cite{Bamba:2014zoa}. Consider a general $f(\phi,R)$ theory coupled to the Gauss-Bonnet invariant $\mathcal{G}$, with the action being equal,
\begin{equation}\label{actionegbgeneral}
\mathcal{S}=\int \mathrm{d}^4x\sqrt{-g}\left(\frac{1}{2}f(\phi,R)-\frac{\omega(\phi )}{2}\partial_{\mu}\phi\partial^{\mu}\phi-V(\phi)-\frac{c_1}{2}\xi (\phi )\mathcal{G} \right)\, ,
\end{equation}
where in our case in the end we take $f(R,\phi)=R$, and $\omega (\phi)=1$, and in addition we assumed a special physical units system, in which $\kappa^2=\frac{8\pi}{M_p}^2=1$, where $M_p$ is the Planck mass scale, and in addition $\hbar=c=1$. Also the Gauss-Bonnet invariant for a general metric $g_{\mu \nu}$ is equal to,
\begin{equation}\label{gaussbonnetscalar}
\mathcal{G}=R^2-4R_{\mu \nu}R^{\mu \nu}+R_{\mu \nu \sigma \rho}R^{\mu \nu \sigma \rho}\, ,
\end{equation}
and for the FRW metric of Eq. (\ref{metricfrw}) it becomes $\mathcal{G}=24 (H^2+\dot{H})$. The action (\ref{actionegbgeneral}) is a subcase of a much more general higher order string inspired Einstein-Gauss-Bonnet Gravity studied in \cite{Nojiri:2006je}, in which case the action is,
\begin{equation}\label{newaction}
\mathcal{S}=\int d^4x\sqrt{-g}\Big{(} \frac{R}{2}-\frac{1}{2}\omega (\phi)\partial_{\mu}\phi\partial^{\mu}\phi-V(\phi )-\xi(\phi)\mathcal{G}-\xi_1(\phi)(R_{\alpha \beta}^{\mu \nu}R^{\alpha \beta}_{\lambda \rho}R^{\mu \nu}_{\lambda \rho}+\epsilon^{\mu \nu \alpha \beta \gamma \delta}\epsilon_{\mu ' \nu '\alpha ' \beta ' \gamma '\delta '}R_{\mu \nu}^{\mu ' \nu '}R_{\alpha \beta}^{\alpha ' \beta '}R_{\gamma \delta}^{\gamma ' \delta '})\Big{)}\, .
\end{equation}
In fact, in certain limits, the above theory can be equivalent to an $F(\mathcal{G})$ theory of gravity \cite{Nojiri:2006je}. By varying the action (\ref{actionegbgeneral}) with respect to the metric $g_{\mu \nu}$ we obtain the following equations of motion for the case of a FRW metric,
\begin{align}\label{eqnsofmotionfrw}
& \frac{\omega (\phi)}{2}\dot{\phi}^2+V(\phi)+\frac{R}{2}F-\frac{f(\phi,R)}{2}-3F(\phi,R)H^2+12c_1\xi'(\phi)\dot{\phi}H^3=0\, ,\\ \notag & \frac{\omega (\phi)}{2}\dot{\phi}^2-V(\phi)+\frac{f(\phi,R)}{2}-3F(\phi,R)(\dot{H}+3H^2)+2\dot{F}H+\ddot{F}-4c_1\left( H^2\dot{\phi}^2\xi''(\phi)+H^2\ddot{\phi}\xi'(\phi)+2H(\dot{H}+H^2)\dot{\phi}\xi'(\phi)\right)=0\, ,
\end{align}
and also by varying with respect to the scalar field we obtain,
\begin{equation}\label{varwiscalar}
\omega (\phi)\ddot{\phi}+3\omega (\phi)H\dot{\phi}V'(\phi)+\frac{1}{2}\omega '(\phi)-\frac{f'(\phi,R)}{2}+12c_1\xi'(\phi)H^2(\dot{H}+H^2)=0\, ,
\end{equation}
where the ``dot'' in the above equations indicates differentiation with respect to the cosmic time, while the ``prime'' indicates differentiation with respect to the scalar field $\phi$, and finally $F=\frac{\partial f(\phi,R)}{\partial R}$. In the minimal case we consider in this paper, we have $f(R,\phi )=R$, $F(\phi,R)=1$ and $f'(\phi,R)=0$.

\subsection{Slow-roll Dynamics}

The dynamics of inflation for generalized Einstein Gauss-Bonnet gravitational theories were studied in Refs. \cite{Noh:2001ia,Hwang:2005hb,Hwang:2002fp,Kaiser:2013sna}, so we shall adopt the notation and formalism of the aforementioned works. The dynamics of inflation are quantified in terms of the slow-roll indices, which in the case of generalized Einstein Gauss-Bonnet gravity appearing in the action (\ref{actionegbgeneral}) are defined as follows,
\begin{equation}\label{slowrollindicesini}
\epsilon_1=\frac{\dot{H}}{H^2},\,\,\,\epsilon_2=\frac{\ddot{\phi}}{H\dot{\phi}},\,\,\,\epsilon_3=\frac{\dot{F}}{2HF},\,\,\,\epsilon_4=\frac{\dot{E}}{2HE},\,\, \,\epsilon_5=\frac{\dot{F}+Q_a}{H(2F+Q_b)},\,\,\,\epsilon_6=\frac{\dot{Q}_t}{2HQ_t}\, ,
\end{equation}
where the function $E$ is defined as follows,
\begin{equation}\label{epsiloncapdefinit}
E=\frac{F}{\dot{\phi}}\left( \omega (\phi)\dot{\phi}^2+3\frac{(\dot{F}+Q_a)^2}{2F+Q_b}\right)\, ,
\end{equation}
and the functions $Q_a$, $Q_b$ and $Q_t$ stand for,
\begin{equation}\label{qaqb}
Q_a=-4c_1\dot{\xi}H^2,\,\,\,Q_b=-8c_1\dot{\xi} H, \,\,\, Q_t=F+\frac{1}{2}Q_b
\end{equation}
The slow-roll approximation for the dynamical inflationary evolution can only be achieved when $\epsilon_i\ll 1$, $i=1,2,..,6$. We shall explicitly calculate the slow-roll indices for the theory at hand, with $\omega=1$ and $f(R,\phi)=R$, and we demonstrate what the slow-roll conditions imply for the functions $\xi (\phi)$ and for the potential $V(\phi)$. By using the first two slow-roll indices, namely $\epsilon_1$ and $\epsilon_2$, the condition on them yields,
\begin{equation}\label{slowrollcond1}
\dot{H}\ll H^2,\,\,\,\ddot{\phi}\ll H\dot{\phi}\, ,
\end{equation}
which are well-known from the canonical scalar field theory. In the case at hand, the slow-roll index $\epsilon_3$ is identically equal to zero, so let us investigate the implications of the slow-roll condition on the slow-roll index $\epsilon_4$. By calculating $\epsilon_4$ we obtain,
\begin{align}\label{epsilon4explicit}
& \epsilon_4=\frac{-\frac{48 c_1^2 H(t)^4 \dot{\xi}(t)^2 \left(-8 c_1 \dot{H}(t) \dot{\xi} (t)-8 c_1 H(t) \ddot{\xi}(t)\right)}{\dot{\phi} (t)^2 \left(2-8 c_1 H(t) \dot{\xi} (t)\right)^2}-\frac{24 c_1 H(t)^2 \dot{\xi} (t) \left(-8 c_1 H(t) \dot{H}(t) \dot{\xi} (t)-4 c_1 H(t)^2 \ddot{\xi} (t)\right)}{\dot{\phi} (t)^2 \left(2-8 c_1 H(t) \dot{\xi} (t)\right)}}{2 H(t) \left(\frac{48 c_1^2 H(t)^4 \dot{\xi} (t)^2}{\dot{\phi} (t)^2 \left(2-8 c_1 H(t) \dot{\xi} (t)\right)}+1\right)},\\ \notag &
-\frac{96 c_1^2 H(t)^4 \dot{\xi}(t)^2 \ddot{\phi} (t)}{\left(\dot{\phi} (t)^3 \left(2-8 c_1 H(t) \dot{\xi} (t)\right)\right) \left(2 H(t) \left(\frac{48 c_1^2 H(t)^4 \dot{\xi} (t)^2}{\dot{\phi} (t)^2 \left(2-8 c_1 H(t) \dot{\xi} (t)\right)}+1\right)\right)}\, ,
\end{align}
and by imposing the condition $\epsilon_4\ll 1$ we obtain the following conditions,
\begin{equation}\label{slowrollcond2}
\dot{\xi}H\ll 1,\,\,\,\ddot{\xi}\ll 1,\,\,\,\dot{\xi}\dot{H}\ll 1\, .
\end{equation}
The slow-roll conditions for the scalar field imply $3H^2\sim V(\phi)$, so by combining the equations of motion (\ref{eqnsofmotionfrw}) and (\ref{varwiscalar}) with the slow-roll conditions, we obtain the following useful relations which yield $\dot{\phi}$ and $\dot{H}$ as functions of the scalar field $\phi$,
\begin{align}\label{derivativesofphiandh}
& \dot{\phi}=-\frac{12c_1\xi'(\phi)V(\phi)^2}{27\sqrt{\frac{V(\phi)}{3}}}-\frac{V'(\phi)}{3\sqrt{\frac{V(\phi)}{3}}}, \\ \notag & \dot{H}=4c_1H^3\xi'(\phi)-\frac{\dot{\phi}^2}{4}\, .
\end{align}
By using the above relations, we obtain the resulting functional form for the slow-roll indices as functions of the scalar field $\phi$, which are,
\begin{align}\label{slowrollindicesasfunctionsofphi}
& \epsilon_1=-\frac{V'(\phi)^2}{4V(\phi)^2},\\ \notag &
\epsilon_2=2\frac{V''(\phi)}{V(\phi)},\\ \notag &
\epsilon_3=0,\\ \notag &
\epsilon_4=-\frac{4 c_1^3 \sqrt{V(\phi )} \xi '(\phi )^3 \left(4 c_1 V(\phi )^2 \xi '(\phi )+3 V'(\phi )\right) \left(20 c_1 V(\phi )^2 \xi '(\phi )+3 V'(\phi )\right)}{27 \sqrt{3}}\, .
\end{align}
In the slow-roll approximation, the spectral index of the primordial curvature perturbations $n_s$ and the tensor-to-scalar ratio $r$ become approximately \cite{Noh:2001ia,Hwang:2005hb,Hwang:2002fp},
\begin{equation}\label{observationalindices1}
n_s\simeq=1+4\epsilon_1-2\epsilon_2+2\epsilon_3-2\epsilon_4,\,\,\,r=4\left(\epsilon_1-\frac{1}{4}(-\frac{Q_e(t)}{H}+Q_f(t)) \right)\, ,
\end{equation}
where the functions $Q_e$ and $Q_f$ are equal to,
\begin{equation}\label{qeqf}
Q_e=8c_1\dot{\xi}\dot{H},\,\,\,Q_f=-4c_1(\ddot{\xi}-\dot{\xi}H)\, .
\end{equation}
By using the slow-roll conditions and also upon combining Eqs. (\ref{derivativesofphiandh}), (\ref{slowrollindicesasfunctionsofphi}), (\ref{observationalindices1}) and (\ref{qeqf}), the observational indices for the Einstein Gauss-Bonnet theory at hand are equal to,
\begin{align}\label{observationalindicesfinalforms}
& n_s\simeq 1-\frac{8 c_1^3 \sqrt{V(\phi )} \xi '(\phi )^3 \left(4 c_1 V(\phi )^2 \xi '(\phi )-3 V'(\phi )\right) \left(4 c_1 V(\phi )^2 \xi '(\phi )+V'(\phi )\right)}{9 \sqrt{3}}-\frac{4 V''(\phi )}{V(\phi )}-\frac{V'(\phi )^2}{V(\phi )^2},\\ \notag & r\simeq \Big{|}-\frac{32}{9} c_1^2 V(\phi )^2 \xi '(\phi )^2-\frac{8}{3} c_1 \xi '(\phi ) V'(\phi )-\frac{2 V'(\phi )^2}{V(\phi )^2}\Big{|}\, .
\end{align}
In the following we shall exemplify our approach by using a well-known class of inflationary potentials, namely power law potentials of the form $V(\phi)\sim \phi^n$. We shall confront the resulting observational indices with the latest Planck data \cite{Ade:2015lrj} and with the BICEP2/Keck-Array data \cite{Array:2015xqh}, and as we will demonstrate, the resulting theory can be compatible with the observations. A vital issue that needs caution, is to see if inflation occurs for large or small values of the scalar field, and also it is important to ensure that during the inflationary era, the slow-roll indices take small values. In the following subsection we shall present in detail the above issues for the aforementioned class of inflationary models.

\subsection{Example I: Power-law Potentials Inflationary Dynamics}

\begin{figure}[h]
\centering
\includegraphics[width=18pc]{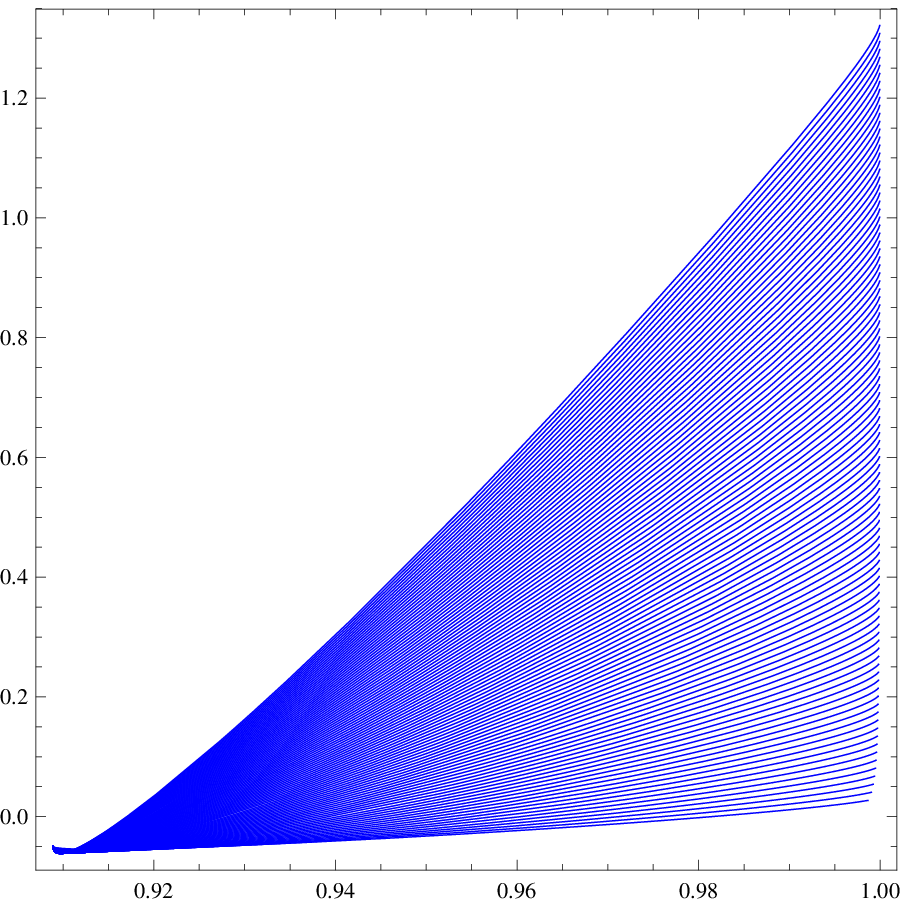}
\includegraphics[width=18pc]{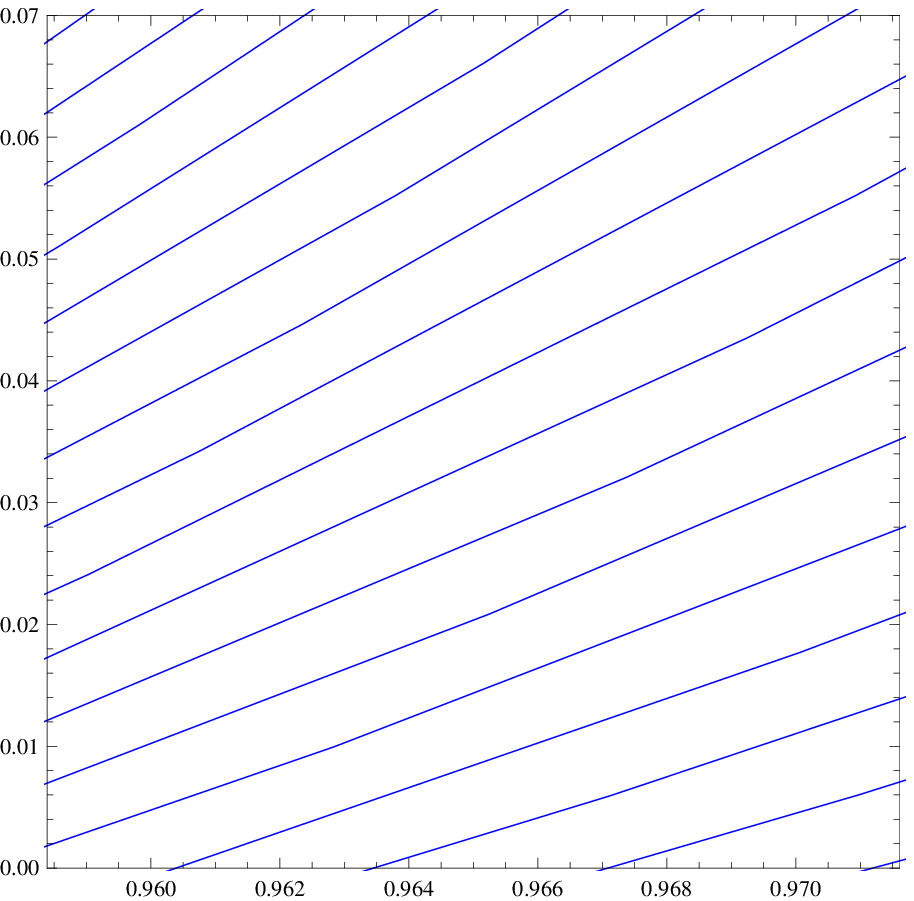}
\caption{The parametric plot of the spectral index (horizontal axis) and of the tensor-to-scalar ratio (vertical axis) for the cubic Einstein Gauss-Bonnet model, as a function of the parameter $\beta$ for $\beta=[0,10]$. The parameter $c_1$ takes values in the range $c_1=[0.5,20]$ with the step being $0.2$. In the right plot $r$ and $n_s$ are in the ranges  $r<0.07$ and $n_s=[0.95839, 0.97161]$, according to the 2018 Planck data and BICEP2/Keck-Array data.}\label{plot1}
\end{figure}

Let us use the results of the previous subsection in order to investigate the viability of a power-law class of potentials of the form $V(\phi)\sim \phi^n$. According to the latest (2015) Planck data , the cubic and quartic potentials are not compatible with the Planck data, so let us investigate whether compatibility with the observations is obtained if the Einstein Gauss-Bonnet theory is used. A similar but quantitatively different approach was adopted in Ref. \cite{Yi:2018gse}. Consider first the cubic potential, so we assume that the potential and the function $\xi (\phi)$ have the following form,
\begin{equation}\label{potentialxithrird}
V(\phi)=V_0\phi^3,\,\,\,\xi (\phi)=-\frac{\beta}{\phi^3}\, ,
\end{equation}
where $V_0$ and $\beta$ are arbitrary dimensionful parameters. The choice of the function $\xi (\phi)$ is arbitrary and in principle any function chosen might be an optimal choice, but we chose it as in Eq. (\ref{potentialxithrird}) in order to obtain simple formulas for the observational indices and for the slow-roll indices as functions of the scalar field $\phi$ and later on as a function of the $e$-foldings number $N$. With the potential $V(\phi)$ and the function $\xi (\phi)$ chosen as in Eq. (\ref{potentialxithrird}), the explicit form of the slow-roll indices as functions of the scalar field are quoted below,
\begin{equation}\label{slowrollindicescase1}
\epsilon_1\simeq -\frac{9}{4 \phi ^2},\,\,\,\epsilon_2\simeq \frac{12}{\phi ^2},\,\,\,\epsilon_4\simeq \frac{4 \sqrt{3} \beta ^3 c_1^3 \sqrt{V_0 \phi ^3} \left(12 \beta  c_1 V_0^2 \phi ^2-9 V_0 \phi ^2\right) \left(12 \beta  c_1 V_0^2 \phi ^2+3 V_0 \phi ^2\right)}{\phi ^{12}}\, ,
\end{equation}
while the corresponding observational indices become,
\begin{align}\label{obsevrationalindciesexample1}
& n_s\simeq 1-\frac{33}{\phi ^2}-\frac{8 \sqrt{3} \beta ^3 c_1^3 \sqrt{V_0 \phi ^3} \left(12 \beta  c_1 V_0^2 \phi ^2-9 V_0 \phi ^2\right) \left(12 \beta  c_1 V_0^2 \phi ^2+3 V_0 \phi ^2\right)}{\phi ^{12}}, \\ \notag &
r\simeq \Big{|}\frac{32 \beta ^2 c_1^2 V_0^2}{\phi ^2}-\frac{24 \beta  c_1 V_0}{\phi ^2}-\frac{18}{\phi ^2}\Big{|}\, .
\end{align}
From the functional form of the slow-roll indices in Eq. (\ref{slowrollindicescase1}), it is obvious that the slow-roll era corresponds to large values of the scalar field $\phi$, so that the condition $\epsilon_i\ll 1$ holds true during the inflationary era. It is worth expressing the observational indices as functions of the $e$-foldings number, which is defined as follows,
\begin{equation}\label{efoldingsdefinition1}
N=\int_{t_i}^{t_f}H(t)\mathrm{d}t\, ,
\end{equation}
where $t_i$, $t_f$ denote the time instances that the inflationary era begins and ends respectively. Expressed in terms of the scalar field $\phi$ in the slow-roll approximation, the $e$-foldings number is written as follows,
\begin{equation}\label{efoldignsscalarfield}
N\simeq \int_{\phi_k}^{\phi_f}-\frac{3 V(\phi )}{4 c_1 V(\phi )^2 \xi '(\phi )+3 V'(\phi )}\, ,
\end{equation}
where $\phi_k$ and $\phi_f$ are the values of the scalar field at the horizon crossing time instance and at the end of the inflationary era respectively. The value $\phi_f$ can be found easily by imposing the condition $|\epsilon_1(\phi_f)|\simeq \mathcal{O}(1)$, so we obtain $\phi_f\simeq \frac{3}{2}$, so indeed the slow-roll inflationary era ends for values of the scalar field $\phi \sim \mathcal{O}(1)$. Accordingly, by using Eq. (\ref{efoldignsscalarfield}), we can find $\phi_k$, which is,
\begin{equation}\label{phikexample1}
\phi_k\simeq \frac{1}{2} \sqrt{32 \beta ^2 c_1 N+24 N+9}\, ,
\end{equation}
so by substituting in Eq. (\ref{obsevrationalindciesexample1}), the observational indices evaluated at the horizon crossing, are approximately equal to,
\begin{equation}\label{nsnexample1}
n_s\simeq 1-\frac{132}{8 N \left(4 \beta ^2 c_1+3\right)+9}-\frac{4608 \sqrt{6} \beta ^3 c_1^3 V_0^2 (4 \beta  c_1 V_0-3) (4 \beta  c_1 V_0+1) \sqrt{V_0 \left(8 N \left(4 \beta ^2 c_1+3\right)+9\right)^{3/2}}}{\left(8 N \left(4 \beta ^2 c_1+3\right)+9\right)^4}\, ,
\end{equation}
\begin{equation}\label{scalartotensorrationexample1}
r\simeq \Big{|} \frac{128 \beta ^2 c_1^2 V_0^2}{32 \beta ^2 c_1 N+24 N+9}-\frac{72}{32 \beta ^2 c_1 N+24 N+9}-\frac{96 \beta  c_1 V_0}{32 \beta ^2 c_1 N+24 N+9} \Big{|}\, .
\end{equation}
Let us now compare the resulting theory with the Planck 2015 \cite{Ade:2015lrj} and BICEP2/Keck-Array data \cite{Array:2015xqh}, which constrain the observational indices as follows,
\begin{equation}
\label{planckdata} n_s=0.9650\pm 0.00661\, , \quad r<0.10\,\,\,(\mathrm{Planck}\,\,\,2018)\, ,
\end{equation}
\begin{equation}
\label{scalartotensorbicep2}
r<0.07\,\,\,(\mathrm{BICEP2}/\mathrm{Keck-Array})\, .
\end{equation}
After a thorough examination of the parameter space, it seems that the Einstein Gauss-Bonnet model at hand is compatible with both the Planck and BICEP2/Keck-Array data, for a large range of values of the free parameters. For example, if $V_0\sim \mathcal{O}(1)$ and $\beta=1.5$, $c_1=0.6$ we get $n_s\simeq 0.9673$ and $r\simeq 0.01$, which are both compatible with the observational constraints (\ref{planckdata}) and (\ref{scalartotensorbicep2}). In order to illustrate this more transparently, in Fig. \ref{plot1} we present the parametric plot of the spectral index $n_s$ and of the tensor-to-scalar ratio $r$, as a function of the parameter $\beta$, for various values of the parameter $c_1$, with $V_0\sim \mathcal{O}(1)$ and for $60$ $e$-foldings. The values of $c_1$ are taken in the range $c_1=[0.5,20]$ with the step being $0.2$ and $\beta$ takes values in the range $\beta=[0,10]$. As it can be seen in Fig. \ref{plot1}, the simultaneous compatibility of $n_s$ and $r$ with the observational data can be achieved for a large range of the free parameters values. Also it is notable that the single canonical scalar field model with  cubic potential without the Gauss-Bonnet coupling yields $n_s\sim 0.9089$ and $r\sim 0.01$, so the spectral index is not compatible with the Planck data. Hence, the Einstein Gauss-Bonnet gravity can make the cubic potential class of models to be compatible with the observations, hence it enhances the phenomenology of otherwise unsuccessful single canonical scalar field models. In the right plot of Fig. \ref{plot1} we plot the spectral index (horizontal axis) and the tensor-to-scalar ratio (vertical axis) for $\beta=[0,10]$ and $c_1=[0.5,20]$, and we constrained $r$ and $n_s$ according to the 2018 Planck data and BICEP2/Keck-Array data (\ref{planckdata}) to be in the range $r<0.07$ and $n_s=[0.95839, 0.97161]$.

\begin{figure}[h]
\centering
\includegraphics[width=18pc]{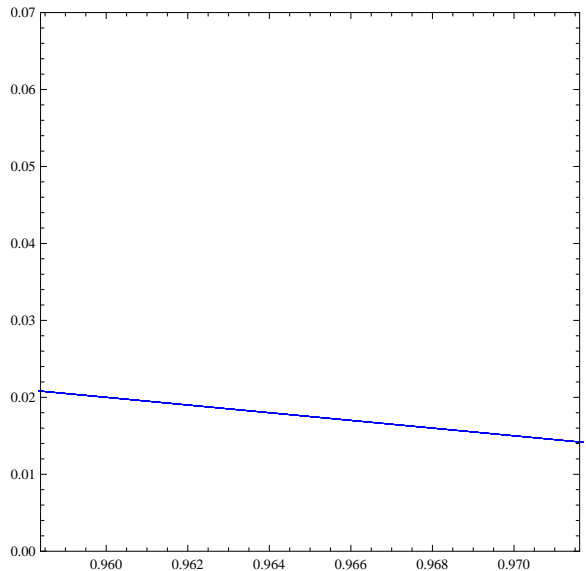}
\caption{The parametric plot of the spectral index (horizontal axis) and of the tensor-to-scalar ratio (vertical axis) for the quartic Einstein Gauss-Bonnet model, as a function of the parameter $\beta$ for $\beta=[0,2]$. The parameter $c_1$ takes values in the range $c_1=[0.5,2]$ with the step being $0.1$. In the plot, $r$ and $n_s$ are in the ranges  $r<0.07$ and $n_s=[0.95839, 0.97161]$, according to the 2018 Planck data.}\label{plot2}
\end{figure}

Let us now consider the case that $V(\phi)$ and $\xi (\phi)$ have the following form,
\begin{equation}\label{potentialxithrirdnew}
V(\phi)=V_0\phi^4,\,\,\,\xi (\phi)=-\frac{\beta}{\phi^4}\, ,
\end{equation}
where $V_0$ and $\beta$ are again arbitrary dimensionful parameters. In this case, the slow-roll indices as functions of the scalar field are equal to,
\begin{equation}\label{slowrollindicescase1new}
\epsilon_1\simeq -\frac{4}{\phi ^2},\,\,\,\epsilon_2\simeq \frac{24}{\phi ^2},\,\,\,\epsilon_4\simeq \frac{256 \beta ^3 c_1^3 \sqrt{V_0 \phi ^4} \left(16 \beta  c_1 V_0^2 \phi ^3-12 V_0 \phi ^3\right) \left(16 \beta  c_1 V_0^2 \phi ^3+4 V_0 \phi ^3\right)}{9 \sqrt{3} \phi ^{15}}\, ,
\end{equation}
while the corresponding observational indices become,
\begin{align}\label{obsevrationalindciesexample1new}
& n_s\simeq 1-\frac{64}{\phi ^2}-\frac{512 \beta ^3 c_1^3 \sqrt{V_0 \phi ^4} \left(16 \beta  c_1 V_0^2 \phi ^3-12 V_0 \phi ^3\right) \left(16 \beta  c_1 V_0^2 \phi ^3+4 V_0 \phi ^3\right)}{9 \sqrt{3} \phi ^{15}},
\\ \notag & r\simeq \Big{|}\frac{512 \beta ^2 c_1^2 V_0^2}{9 \phi ^2}-\frac{128 \beta  c_1 V_0}{3 \phi ^2}-\frac{32}{\phi ^2}\Big{|}\, .
\end{align}
By looking the slow-roll indices in Eq. (\ref{slowrollindicescase1new}), it is obvious that in this case too, the slow-roll inflationary era is obtained when the scalar field takes large values. By following the procedure of the previous case, the observational indices as functions of the $e$-foldings number read,
\begin{equation}\label{nsnexample1new}
n_s\simeq 1-\frac{48}{8 \beta ^2 c_1 N+6 N+3}-\frac{192 \beta ^3 c_1^3 (4 \beta  c_1 V_0-3) (4 \beta  c_1 V_0+1) \left(V_0 \left(N \left(8 \beta ^2 c_1+6\right)+3\right)^2\right)^{5/2}}{\left(N \left(8 \beta ^2 c_1+6\right)+3\right)^{17/2}}\, ,
\end{equation}
\begin{equation}\label{scalartotensorrationexample1new}
r\simeq \Big{|} \frac{128 \beta ^2 c_1^2 V_0^2}{3 \left(8 \beta ^2 c_1 N+6 N+3\right)}-\frac{24}{8 \beta ^2 c_1 N+6 N+3}-\frac{32 \beta  c_1 V_0}{8 \beta ^2 c_1 N+6 N+3} \Big{|}\, .
\end{equation}
The compatibility with the observational data can be achieved for the quartic model, however for narrower range of parameter values. For example by choosing $V_0\sim \mathcal{O}(10^{-4})$ and for $\beta=2.05$ and $c_1=0.5$ we get $n_s\simeq 0.965004$ and $r\simeq 0.0175002$, which are both compatible with the observational constraints (\ref{planckdata}) and (\ref{scalartotensorbicep2}). In Fig. \ref{plot2} we have presented the parametric plot of the spectral index $n_s$ and of the tensor-to-scalar ratio $r$, as functions of the parameter $\beta$ with $V_0\sim \mathcal{O}(10^{-4})$ and for $60$ $e$-foldings. The values of $c_1$ are $c_1=[0.5,2]$ with the step being $0.1$ and $\beta$ takes values in the range $\beta=[0,2]$. As it can be seen in Fig. \ref{plot2}, a simultaneous compatibility of the observational indices $n_s$ and $r$ can be achieved, but for a narrow range of parameter values. In the plot, the observational indices $r$ and $n_s$ are assumed to take values in the ranges  $r<0.07$ and $n_s=[0.95839, 0.97161]$, according to the 2018 Planck data and BICEP2/Keck-Array data. However the single canonical scalar field theory with a quartic potential yields $n_s\simeq 0.867769$ and $r\simeq 0.0661157$ for $60$ $e$-foldings, so the spectral index of the corresponding canonical scalar field theory is excluded by the observational data. Hence, the Einstein Gauss-Bonnet theory modification of the quartic scalar field theory enhances the phenomenological viability of the latter.

\subsection{Example II: Simple Exponential Potentials Inflationary Dynamics}

\begin{figure}[h]
\centering
\includegraphics[width=18pc]{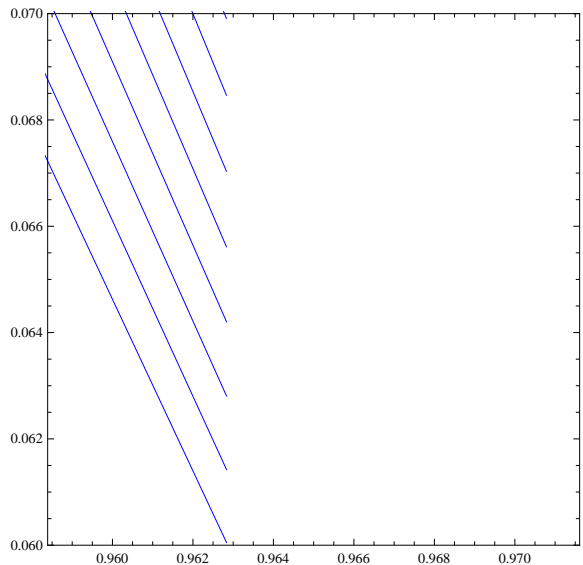}
\caption{The parametric plot of the spectral index $n_s$ and of the tensor-to-scalar ratio $r$, as functions of the parameters $q$ and $V_0$ for  $N=60$ $e$-foldings, $D_0=150$ and $c_1=1$, for the Einstein Gauss-Bonnet exponential model with $V(\phi)=V_0e^{-q\phi}$ and $\xi (\phi)=D_0e^{q\phi}$. The values of $q$ are chosen in the range $q=[1/10,1/11.6]$ and $V_0=[0.009,0.01]$ with step $0.001$. In the plot, $r$ and $n_s$ are in the ranges  $r<0.07$ and $n_s=[0.95839, 0.97161]$, according to the 2018 Planck data.}\label{plot2anew}
\end{figure}

Let us now consider another appealing example in which the Einstein Gauss-Bonnet extension of the canonical scalar field theory may provide an appealing solution to a major theoretical problem of the single scalar theory. Particularly we shall consider the case for which the scalar field potential $V(\phi)$ and the scalar coupling function $\xi (\phi)$ have the following form \cite{Nojiri:2005vv},
\begin{equation}\label{potentialxithrirdnew}
V(\phi)=V_0e^{-q\phi},\,\,\,\xi (\phi)=D_0e^{q\phi}\, ,
\end{equation}
where $V_0$, $q$ and $D_0$ are again positive arbitrary dimensionful parameters. The canonical scalar theory can be compatible with the observational data, however the classical single scalar theory is haunted by the graceful exit issue. Particularly, the classical theory has constant slow-roll indices, and although it can be arranged to have small values, so that a slow-roll era can be realized, these can never become of order $\mathcal{O}(1)$, and hence there is now actual mechanism to trigger the graceful exit from inflation. This issue of graceful exit was also discussed in \cite{Kawai:1999pw,Antoniadis:1993jc,Mavromatos:2000az,Hernandez:2002ig,Gasperini:1996fu,Brustein:1997cv,Easson:1999xw,Cartier:1999vk} some time ago. As we will explicitly demonstrate in a quantitative way, the Einstein Gauss-Bonnet theory can provide a mechanism that may trigger graceful exit from inflation via the slow-roll index $\epsilon_4$ which is field-dependent and it can trigger the exit from inflation when it becomes of the order one. Let us demonstrate this by calculating the slow-roll indices, which are,
\begin{equation}\label{slowrollindicescase1newnew}
\epsilon_1\simeq -\frac{q^2}{4},\,\,\,\epsilon_2\simeq 2 q^2,\,\,\,\epsilon_4\simeq \frac{4 c_1^3 D_0^3 q^5 V_0^3 (4 c_1 D_0 V_0-1) (4 c_1 D_0 V_0+3)}{9 \sqrt{3} \sqrt{V_0 e^{-q \phi }}}\, ,
\end{equation}
and the corresponding observational indices become in this case,
\begin{align}\label{obsevrationalindciesexample1newnew}
& n_s\simeq 1-\frac{64}{\phi ^2}-\frac{512 \beta ^3 c_1^3 \sqrt{V_0 \phi ^4} \left(16 \beta  c_1 V_0^2 \phi ^3-12 V_0 \phi ^3\right) \left(16 \beta  c_1 V_0^2 \phi ^3+4 V_0 \phi ^3\right)}{9 \sqrt{3} \phi ^{15}},
\\ \notag & r\simeq \Big{|}\frac{512 \beta ^2 c_1^2 V_0^2}{9 \phi ^2}-\frac{128 \beta  c_1 V_0}{3 \phi ^2}-\frac{32}{\phi ^2}\Big{|}\, .
\end{align}
By looking the slow-roll indices in Eq. (\ref{slowrollindicescase1newnew}), it can be seen that the first two slow-roll indices are constant, but the slow-roll $\epsilon_4$ is field dependent, thus it can potentially be of the order one for some field value. By requiring $\epsilon_4=1$ we obtain the value of the scalar field when this condition holds true, which we denote as $\phi_f$ and it is equal to,
\begin{equation}\label{phifinal}
\phi_f=\frac{\log \left(\frac{16}{243} c_1^6 D_0^6 q^{10} V_0^5 \left(16 c_1^2 D_0^2 V_0^2+8 c_1 D_0 V_0-3\right)^2\right)}{q}\, .
\end{equation}
Accordingly, by using the method of the previous section, the value of the scalar field at the horizon crossing, as a function of the $e$-foldings is equal to,
\begin{equation}\label{phihorizoncrossexponential}
\phi_k=\frac{\log \left(\frac{16}{243} c_1^6 D_0^6 q^{10} V_0^5 \left(16 c_1^2 D_0^2 V_0^2+8 c_1 D_0 V_0-3\right)^2\right)}{q}+\frac{1}{3} N q \left(4 c_1 V_0^2-3\right)\, .
\end{equation}
Hence, by calculating the observational indices at the horizon crossing value $\phi=\phi_k$, these become equal to,
\begin{align}\label{nsnexample1newnew}
& n_s\simeq -\frac{1}{243} 32 c_1^9 D_0^9 q^{15} V_0^7 (4 c_1 D_0 V_0-1)^3 e^{\frac{1}{3} N q^2 \left(4 c_1 V_0^2-3\right)}-5 q^2+1, \\ \notag &
\times (4 c_1 D_0 V_0+3)^3 \sqrt{\frac{e^{-\frac{1}{3} N q^2 \left(4 c_1 V_0^2-3\right)}}{c_1^6 D_0^6 q^{10} V_0^4 \left(16 c_1^2 D_0^2 V_0^2+8 c_1 D_0 V_0-3\right)^2}}\, ,
\end{align}
\begin{equation}\label{scalartotensorrationexample1newnew}
r\simeq \Big{|} \frac{32}{9} c_1^2 D_0^2 q^2 V_0^2+\frac{8}{3} c_1 D_0 q^2 V_0-2 q^2 \Big{|}\, .
\end{equation}
The above observational indices can easily become compatible with the observational data for a wide range of the free parameters values. For example, by choosing $V_0=0.001$, $q=1/12.5$, $D_0=150$ and $c_1=1$ for 60 $e$-foldings, we obtain the following values,
\begin{equation}\label{exponentialobservresults}
n_s\simeq 0.968,\,\,\,r\simeq 0.009728\, ,
\end{equation}
which are compatible with the observational constraints coming from Planck (\ref{planckdata}) and the BICEP2/Keck Array data (\ref{scalartotensorbicep2}). In order to further demonstrate the viability of the resulting theory for various parameter values, in Fig. \ref{plot2anew} we present the parametric plot of the spectral index $n_s$ and of the tensor-to-scalar ratio $r$, as functions of the parameters $q$ and $V_0$ by choosing  $N=60$ $e$-foldings, and for $D_0=150$ and $c_1=1$. The values of $q$ are chosen in the range $q=[1/10,1/11.6]$ and $V_0=[0.009,0.01]$ with step $0.001$. The curves in the parametric plot indicate the parameter values for which $n_s$ and $r$ are simultaneously compatible with the observational data.
From the above results we conclude that the resulting exponential Einstein-Gauss Bonnet exponential inflation model is compatible with the Planck and the BICEP2/Keck-Array data, and more importantly, it also has an inherent mechanism to trigger the graceful exit from inflation. Hence the viability and physical significance of the canonical single scalar theory is enhanced by the Einstein Gauss-Bonnet inflation theory.

\section{Reconstruction of Viable Einstein Gauss-Bonnet Inflation from the Observational Indices}

In the previous section we adopted a direct technique to study inflationary dynamics, namely we chose the potential and the scalar function $\xi (\phi)$, and by using these we found the observational indices in the slow-roll approximation. In this section we shall adopt a bottom-up approach in order to realize viable inflationary scenarios in the context of Einstein Gauss-Bonnet gravity. In some previous works \cite{Odintsov:2018ggm,Odintsov:2017fnc} we have applied this bottom-up approach in various modified gravities, for example $f(R)$ gravity \cite{Odintsov:2017fnc} and $f(R,\phi)$ gravity \cite{Odintsov:2018ggm}. The bottom-up approach has the privilege of starting with a viable theory, and by using the observational indices and specifically the tensor-to-scalar ratio, one seeks for the theory that may realize this phenomenology. In the following we shall assume that the slow-roll conditions hold true, in order to simplify the reconstruction procedure.

\begin{figure}[h]
\centering
\includegraphics[width=18pc]{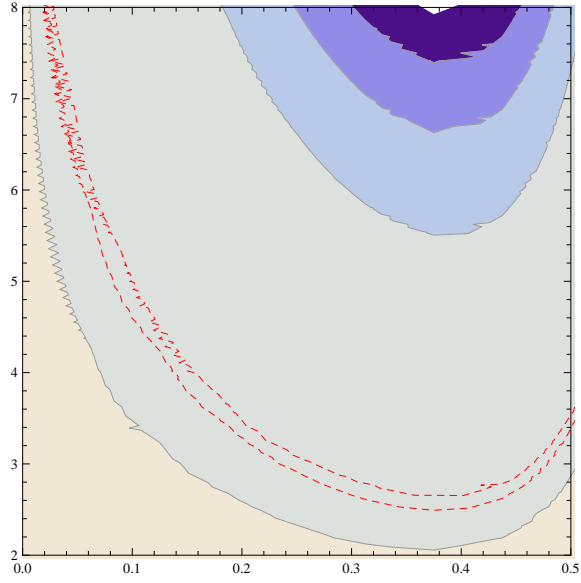}
\caption{The spectral index  for the quartic Einstein Gauss-Bonnet model, as a function of the parameters $\beta$ and $\gamma$ for $N=60$ and $\delta=2.3$. The parameters  $\beta$ and $\gamma$ take values in the ranges $\beta=[0,8]$ and $\gamma=[0,0.5]$. The red dashed curves correspond to the values $n_s=0.97161$ and $n_s=0.95839$, and the region between these two curves denotes the allowed values of $n_s$ according to the 2018 Planck data. The lowest curve in the graph corresponds to the value $n_s=1$.}\label{plot3}
\end{figure}

The bottom-up approach we shall adopt is based on using the $e$-foldings number $N$ as the dynamical variable instead of the cosmic time. We can easily express the various physical quantities of the theory in terms of the $e$-foldings number $N$, by using the following differentiation rules,
\begin{equation}\label{trick1}
\frac{\mathrm{d}}{\mathrm{d}t}=H\frac{\mathrm{d}}{\mathrm{d}N}\, ,
\end{equation}
\begin{equation}\label{asxet1}
\frac{\mathrm{d}^2}{\mathrm{d}t^2}=H^2\frac{\mathrm{d}^2}{\mathrm{d}N^2}+H\frac{\mathrm{d}H}{\mathrm{d}N}\frac{\mathrm{d}}{\mathrm{d}N}\,
.
\end{equation}
Then, after some algebra it can be shown that the slow-roll indices defined in Eq. (\ref{slowrollindicesini}), can be written as follows,
\begin{align}\label{slowrollindicesreconemeth}
& \epsilon_1=\frac{H'(N)}{H(N)},\\ \notag &
\epsilon_2=\frac{H(N) H'(N) \phi '(N)+H(N)^2 \phi ''(N)}{H(N)^2 \phi '(N)},\\ \notag &
\epsilon_3=0,\\ \notag &
\epsilon_4=48 c_1^3 H(N)^9 H'(N) \xi '(N)^3 \phi '(N)\, ,
\end{align}
where the ``prime'' in the equation above and for the rest of this section, denotes differentiation with respect to the $e$-foldings number $N$. Accordingly, the tensor-to-scalar ratio is equal to,
\begin{equation}\label{tensortoscalarratind}
r=\frac{1}{2} c_1 H(N)^2 \xi '(N)+\frac{4 H'(N)}{H(N)}\, .
\end{equation}
The reconstruction method is based on the following steps, firstly one should define the Hubble rate as a function of the $e$-foldings number. Then the tensor-to-scalar ratio is assumed to have an appropriate functional form, so that it corresponds to a viable phenomenology. Then by substituting $H(N)$ and $r$ in Eq. (\ref{tensortoscalarratind}) we can solve the resulting differential equation with respect to $\xi (N)$. Having $\xi (N)$ at hand enables us to find the function $\phi (N)$ and also the scalar potential $V(N)$, and in effect to find the explicit form of the spectral index of primordial curvature perturbations. Let us see how these steps can be realized, so firstly from the equations of motion upon combining the first two appearing in Eq. (\ref{eqnsofmotionfrw}), after some algebra we get,
\begin{equation}\label{phi3asfn}
\left(\phi'(N)\right)^{3}\simeq 8c_1H(N)\xi'(N)\, ,
\end{equation}
and hence by solving the above differential equation, the approximate explicit form of the function $\phi (N)$ is obtained. Substituting $\phi (N)$ in Eq. (\ref{varwiscalar}) and by neglecting $\ddot{\phi}$ we get the following differential equation,
\begin{equation}\label{diffeqnforVn}
\frac{V'(N)}{\phi'(N)}=-3H(N)^2\phi'(N)+\frac{12c_1\xi'(N)H(N)^4}{\phi'(N)}\, ,
\end{equation}
which can be solved with respect to $V(N)$ and therefore the potential as a function of the $e$-foldings is also obtained. Having $\xi (N)$, $\phi (N)$ and $V(N)$ at hand, one can easily obtain the spectral index of the primordial curvature perturbations as a function of the $e$-foldings, and its explicit form is given below,
\begin{equation}\label{nsbottomupreconstruction}
n_s\simeq 1+\frac{2 H'(N)}{H(N)}-96 c_1^3 H(N)^9 H'(N) \xi '(N)^3 \phi '(N)-\frac{2 \phi ''(N)}{\phi '(N)}\, .
\end{equation}
Hence, one can obtain a viable theory by appropriately choosing the Hubble rate and the tensor-to-scalar ratio, having always in mind that the chosen Hubble rate must describe an inflationary evolution. Also, if the function $\phi (N)$ can be inverted, the explicit forms of the functions $\xi (\phi)$ and $V(\phi)$ can be found. It is worth exemplifying the bottom-up reconstruction technique, so let us demonstrate how this technique functions, by choosing the Hubble rate to be,
\begin{equation}\label{hubbleratebottomuprecon}
H(N)=\beta  N^{\gamma }\, ,
\end{equation}
where $\beta$ and $\gamma$ are real positive numbers. Also we assume that the tensor-to-scalar ratio has the following form,
\begin{equation}\label{ternsortoscalarexplicitformburt}
r=\frac{\delta}{N}\, ,
\end{equation}
where $\delta$ is some real positive number. The allowed values of the parameter $\delta$ can be constrained from the BICEP2/Keck-Array data, so since the tensor-to-scalar ratio is constrained to satisfy $r<0.07$, for 60 $e$-foldings this means that $\delta<4.2$. We shall take this constraint on $\delta$ into account later on in this section. By substituting Eqs. (\ref{hubbleratebottomuprecon}) and (\ref{ternsortoscalarexplicitformburt}) in Eq. (\ref{tensortoscalarratind}), we obtain the following differential equation,
\begin{equation}\label{diffeqnforxn}
\frac{1}{2} \beta ^2 c_1 N^{2 \gamma } \xi '(N)+\frac{4 \gamma }{N}=\frac{\delta}{N}\, ,
\end{equation}
and by solving it with respect to $\xi (N)$, we obtain,
\begin{equation}\label{xiNsolution}
\xi (N)=\Lambda-\frac{(2 \delta -8 \gamma ) N^{-2 \gamma }}{2 \beta ^2 \gamma c_1}\, ,
\end{equation}
where $\Lambda$ is an integration constant. By substituting the resulting $\xi (N)$ in Eq. (\ref{phi3asfn}) we obtain $\phi (N)$ which reads,
\begin{equation}\label{phinexplicit1}
\phi (N)\simeq \Lambda -\frac{6 \sqrt{3}{2} \sqrt{3}{\delta -4 \gamma } N^{\frac{1}{3} (-\gamma -1)+1}}{\sqrt[3]{\beta } (\gamma -2)}\, .
\end{equation}
Accordingly, by substituting the resulting $\phi (N)$ and $\xi (N)$ from Eqs. (\ref{xiNsolution}) and (\ref{phinexplicit1}) in the differential equation (\ref{diffeqnforVn}), we obtain the function $V(N)$ which reads,
\begin{equation}\label{vnexampleq1}
V(N)\simeq -\frac{36\ 2^{2/3} \beta ^{4/3} (\delta -4 \gamma )^{2/3} N^{\frac{4 \gamma }{3}+\frac{1}{3}}}{4 \gamma +1}-\frac{12 \beta ^2 (4 \gamma -\delta ) N^{2 \gamma }}{\gamma }+\Lambda_1\, ,
\end{equation}
where $\Lambda_1$ is an integration constant. Finally, by substituting Eqs. (\ref{xiNsolution}), (\ref{phinexplicit1}) and (\ref{vnexampleq1}) in Eq. (\ref{nsbottomupreconstruction}), we obtain the spectral index of the primordial curvature perturbations for the gravitational theory at hand, as a function of the $e$-foldings number, which is,
\begin{equation}\label{nsasfunctionofN}
n_s\simeq \frac{(8 \gamma +2) N^{10/3}+3 N^{13/3}-4608 \sqrt[3]{2} \beta ^{11/3} \gamma  (\delta -4 \gamma )^{10/3} N^{\frac{11 \gamma }{3}}}{3 N^{13/3}}\, .
\end{equation}
With the spectral index (\ref{nsasfunctionofN}) and the tensor-to-scalar ratio at hand, the viability of the resulting inflationary theory can be explicitly checked. Indeed, by choosing for example $\beta \sim \mathcal{O}(1)$ and $\delta=3.6$, $\gamma =0.4$ we obtain $n_s\simeq 0.966535$ and $r=0.06$ which are compatible with the 2018 Planck (\ref{planckdata}) and BICEP2/Keck-Array constraints (\ref{scalartotensorbicep2}). The compatibility with the observational data can be achieved for a considerable range of parameter values, and in order to demonstrate this in Fig. \ref{plot3} we present the contour plot of the spectral index $n_s$ for $N=60$ and $\delta=2.3$. The tensor-to-scalar ratio for these values of $N$ and $\delta$ reads $r\simeq 0.038$, so it is compatible with the BICEP2/Keck-Array data. As it can be seen in Fig. \ref{plot3} the region between the two red and dashed curves leads to a phenomenologically viable spectral index, since these two curves indicate the values $n_s=0.0.97161$ and $n_s=0.95839$, which are the upper and lower bound of the Planck data constraints on $n_s$. The parameters  $\beta$ and $\gamma$ take values in the ranges $\beta=[0,8]$ and $\gamma=[0,0.5]$.
Finally, it is also possible for this case to find the explicit form of the potential $V(\phi)$ and of the function $\xi (\phi)$ by inverting the function $\phi (N)$ appearing in Eq. (\ref{phinexplicit1}), and by doing so and substituting the result in Eq. (\ref{vnexampleq1}), the scalar potential reads,
\begin{align}\label{vpjhiexplicitform}
& V(\phi)\simeq -\frac{2^{\frac{4 (4 \gamma +1)}{3 (\gamma -2)}+\frac{8}{3}} 3^{\frac{4 \gamma +1}{\gamma -2}+2} (\gamma -2)^{-\frac{4 \gamma +1}{\gamma -2}} \beta ^{\frac{5}{3}-\frac{4 \gamma +1}{\gamma -2}} \sqrt[3]{\delta -4 \gamma } (\Lambda -\phi )^{-\frac{4 \gamma +1}{\gamma -2}}}{4 \gamma +1}\\ \notag & -\frac{3^{\frac{3}{\gamma -2}+1} 4^{\frac{2}{\gamma -2}+1} (\gamma -2)^{-\frac{6 \gamma }{\gamma -2}} \beta ^{\frac{7}{3}-\frac{6 \gamma }{\gamma -2}} (4 \gamma -\delta ) (\Lambda -\phi )^{-\frac{6 \gamma }{\gamma -2}}}{\gamma  \sqrt[3]{\delta -4 \gamma }}+\Lambda_1\, .
\end{align}
Accordingly by substituting  $N(\phi )$ in Eq. (\ref{xiNsolution}) we obtain the function $\xi (\phi)$, which is,
\begin{equation}\label{xiphiscalarfphiform}
\xi (\phi)\simeq \frac{2^{-\frac{8 \gamma }{\gamma -2}-1} 3^{-\frac{6 \gamma }{\gamma -2}} (2-\gamma )^{\frac{6 \gamma }{\gamma -2}} \beta ^{\frac{6 \gamma }{\gamma -2}-\frac{5}{3}} (2 \delta -8 \gamma ) (\Lambda -\phi )^{\frac{6 \gamma }{\gamma -2}}}{\gamma  c_1 \sqrt[3]{\delta -4 \gamma }}+\Lambda\, .
\end{equation}
With the bottom-up reconstruction technique we presented in this section, it is possible in principle to realize various classes of viable inflationary models, however for brevity we confined ourselves to a relatively calculational easy and concrete example.

\subsection{Stability Issues of the Slow-roll Theory}

An important issue we did not address so far in this paper in the stability issues of the theory at hand. This issue was thoroughly addressed in Ref. \cite{Calcagni:2006ye} for a wider class of theories, in which the ones we studied in this paper are some subclass. The authors in \cite{Calcagni:2006ye} investigated the no ghost conditions for tensor, vector and scalar perturbations and in addition they investigated the conditions that prevent superluminal propagation of perturbations modes. Using the notation of Ref. \cite{Calcagni:2006ye}, and combining with our notation, we have the following correspondence:
\begin{equation}\label{mnotation}
\omega=1,\,\,\,\alpha_4=0,\,\,\,f_2(\phi)=-\frac{c_1}{2}\xi (\phi ),\,\,\,f_1(\phi)=1,\,\,\,\gamma_{11}=1,\,\,\,\gamma_{12}=0\, .
\end{equation}
Hence let us investigate the various constraints imposed for cosmological perturbations from Ref. \cite{Calcagni:2006ye}. We start off with the tensor perturbations, in which case we have the following two constraints:
\begin{align}\label{constraints1}
& 1+8 \kappa^2 H \dot{f}_2>0\\ \notag &
1+8 \kappa^2 \ddot{f}_2>0\, ,
\end{align}
while for the scalar perturbations, since in our case $\omega \geq 1$ and $\alpha_4=0$, no extra condition apart from the ones above, are imposed in the theory. Finally, with regard to the superluminal propagation of perturbation modes, in the case of $\alpha_4=0$, which is our case, the superluminal constraint is $\dot{f}_2>0$. Hence, the constraints (\ref{constraints1}) along with $\dot{f}_2>0$ should also be taken into account and these may restrict the allowed parameter space. However, the examination of these conditions is an analytically non-trivial task, and this issue should be addressed numerically in the context of the slow-roll approximation, so we defer this study in a detailed and focused on this issue work.

\section{Conclusions}

In this paper the focus was on inflationary solutions of Einstein Gauss-Bonnet gravity. Particularly we investigated the implications of the slow-roll conditions on the gravitational theory by imposing the condition that the slow-roll indices are much smaller than unity during the slow-roll era. In turn the slow-roll condition restricted the gravitational theory quantified by the coupling of the Gauss-Bonnet scalar $\xi (\phi)$ and by the scalar potential $V(\phi)$. After discussing the general formalism, we investigated the phenomenological implications of power-law and exponential potentials focusing on the cubic, quadratic order potentials and simple exponential potentials. The power-law potentials in the context of a single canonical scalar field theory fail to comply with the Planck 2015 observational data, however as we demonstrated, in the context of Einstein Gauss-Bonnet theory it is possible to obtain a phenomenologically viable theory, and in fact for a wide range of values of the free parameters of the theory. Hence a first outcome of this work is that the Einstein Gauss-Bonnet theories can make non-viable single canonical scalar field theories to be viable, and actually the coupling of the scalar field to the Gauss-Bonnet scalar controls this procedure. With regard to the exponential potentials, the classical single scalar theory has no inherent mechanism to trigger the graceful exit from inflation, since the slow-roll indices are constant and field-independent. However the Einstein Gauss-Bonnet version of the theory has the slow-roll index $\epsilon_4$ which is field dependent, and thus the slow-roll phase ends when this index becomes of order $\mathcal{O}(1)$. In the second part of this work we presented a bottom-up reconstruction method in the context of which it is possible to obtain a viable inflationary theory by fixing the functional form of the observational indices, and particularly of the tensor-to-scalar ratio. Specifically, one fixes the Hubble rate and the tensor-to-scalar ratio as functions of the $e$-foldings number. After that, by appropriately expressing the slow-roll indices and the equations of motion as functions of the $e$-foldings number, it is possible to obtain the exact Einstein Gauss-Bonnet theory that may realize the given Hubble rate and the tensor-to-scalar ratio, provided that the slow-roll conditions hold true. Particularly, one obtains the functions $\xi (\phi (N)))$, and $\phi (N)$, and from these the potential $V(N)$ is obtained. By using the resulting functions, the calculation of the spectral index of the primordial curvature perturbations is straightforward, and hence it can be checked when the resulting theory is compatible with the observational data. Finally, if the resulting function $\phi (N)$ can be inverted, the scalar potential $V(\phi)$ and the function $\xi (\phi)$ can be obtained.

In principle, the methodology we introduced in this work can be applied in generalized forms of Einstein Gauss-Bonnet gravity, in which case the gravitational part containing the scalar curvature $R$ may be a function of $R$, that is, an $f(R)$ gravity, or it may be a function of both the scalar curvature and of the scalar field of the form $f(R,\phi)$. In this case the resulting equations of motion might be more complicated, however the slow-roll conditions on the indices may simplify the calculations to a great extent. We hope to address soon some of the above issues in a future work.

\section*{Acknowledgments}

This work is supported by MINECO (Spain), FIS2016-76363-P (S.D.O) and by PHAROS-COST action No: CA16214 (S.D. Odintsov and V.K. Oikonomou).

\end{document}